\documentclass{mem}
\usepackage{natbib}\usepackage{txfonts}\usepackage{balance}
\usepackage{graphicx}
\usepackage[a4paper]{hyperref}
\idline{75}{282}
\begin{document}

\title{
Observational results for northern and southern (candidate) $\gamma$ Doradus
stars   
}

\subtitle{}

\author{
P.~De Cat\inst{1,2} \and K.~Goossens\inst{2} \and F.~Bouckaert\inst{2} \and
L.~Eyer\inst{2,3} \and J.~Cuypers\inst{1} \and J.~De
Ridder\inst{2,}\thanks{Postdoctoral Fellow of the Fund for Scientific
  Research, Flanders} \and C.~Aerts\inst{2,4} \and M.-A.~Dupret\inst{5} \and
A.~Grigahc\`ene\inst{6} \and many observers          
}

\offprints{peter@oma.be}
 
\institute{
Royal Observatory of Belgium, Ringlaan 3, B-1180 Brussel, Belgium
\and
Instituut voor Sterrenkunde, Katholieke Universiteit Leuven, Celestijnenlaan
200 B, B-3001 Leuven, Belgium
\and
Observatoire de Gen\`eve, CH-1290 Sauverny, Switzerland
\and 
Department of Astrophysics, University of Nijmegen, PO Box 9010, 6500 GL
Nijmegen, the Netherlands
\and
Observatoire de Paris, LESIA, 92195 Meudon, France
\and
CRAAG - Algiers Observatory BP 63 Bouzareah 16340, Algiers, Algeria
}

\authorrunning{De Cat et al. }

\titlerunning{Observational results for (candidate) $\gamma$ Doradus stars}

\abstract{
We report on observational results obtained for 78 objects originally
classified as bona-fide or candidate $\gamma$ Doradus stars.
For the southern objects, we gathered echelle spectra with the {\sc
  coralie} spectrograph attached to the Euler telescope in 1998--2003
and/or Johnson-Cousins $B,V,I_c$ observations with the {\sc modular} photometer attached
to the 0.5-m {\sc saao} telescope in 1999--2000.
For the northern objects, we obtained Geneva $U,B,B_1,B_2,V,V_1,G$
observations with the {\sc p7} photometer attached to the 1.2-m Mercator
telescope in 2001--2004.
At least 15 of our objects are binaries, of which 7 are new. 
For 6 binaries, we determined the orbit for the first time.
At least 17 objects show profile variations and at least 12 objects are
multiperiodic photometric variables.
Our results allow us to upgrade 11 objects to bona-fide $\gamma$ Doradus stars
and to downgrade 8 objects to constant up to the current detection limits.
Mode identification is still ongoing, but so far, only $\ell$\,= 1 and 2 modes
have been identified.
}
\maketitle{}

\section{Introduction} \label{intro}

$\gamma$\,Doradus ($\gamma$\,Dor) stars are variable late A- to early F-type stars situated along
the main-sequence. 
Their observed multi-periodic variations with typical periods between 0.5 and
3 days are attributed to non-radial $g$-modes driven by a flux blocking
mechanism at the base of the convective envelope
(e.g. \citealt{Dupret2004A&A...414L..17D}).  
Currently, there are 54 bona-fide, 104 candidate and 15 rejected $\gamma$\,Dor stars
known
\citep{Handler2002,Henry2002PASP..114..988H,Henry2003AJ....126.3058H,Fekel2003AJ....125.2156F,Mathias2004A&A...417..189M,Henry2005AJ....129.2026H,Henry2005AJ....129.2815H}. 
We contribute to the observational effort which is made in recent years to
classify $\gamma$\, Doradus stars in order to further
constrain both their characteristics and their instability strip in the HR
diagram.

\section{Data} \label{data}

For a sample of 36 southern objects, we gathered a total of 620 echelle spectra
during 28 weeks in 1998--2003 with the {\sc coralie} spectrograph attached to the
1.2-m Euler telescope (La Silla, Chile).
We cross-correlated them with the standard template of an F0-type star to
derive accurate radial velocities. 
For another sample of 37 southern objects, of which 28 are in common with the
spectroscopic survey, we obtained a total of 3913 photometric observations in
the Johnson-Cousins $B,V,I_c$ filters during 15 weeks in 1999--2000 with the 0.5-m
{\sc SAAO} telescope (Sutherland, South-Africa). 
For a sample of 36 northern objects, we obtained a total of 3878 photometric
observations in the Geneva $U,B,B_1,B_2,V,V_1,G$ filters during more than 80
weeks in 2001--2004 with the 1.2-m {\sc mercator} telescope (La Palma, Spain). 

\section{Orbital variations} \label{orbital}

\begin{figure}[]
\begin{center}
\resizebox{0.75\hsize}{!}{\rotatebox{270}{\includegraphics[clip=true]{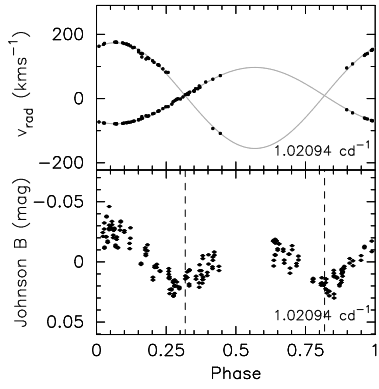}}}
\caption{
\footnotesize 
Phase diagram of the radial velocity $v_{\rm rad}$ (top) and the Johnson $B$
measurements (bottom) of HD\,81421 with the orbital frequency as given in the
bottom right corner. 
The reference epoch is HJD 2450000.
The dashed lines in the bottom left panel denote the phase at which $v_{\rm rad}$\,=
$v_{\gamma}$.
}
\label{81421}
\end{center}
\end{figure}

\begin{figure}[]
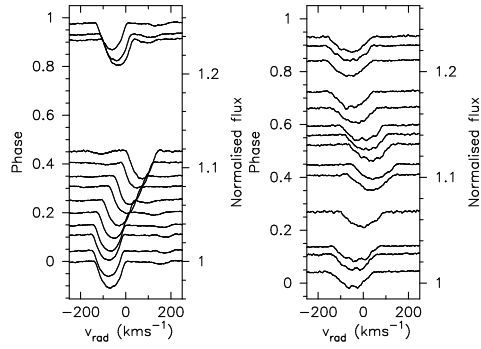

\begin{center}
\resizebox{0.47\hsize}{!}{\rotatebox{270}{\includegraphics[clip=true]{DeCatfig2a.ps}}}
\resizebox{0.47\hsize}{!}{\rotatebox{270}{\includegraphics[clip=true]{DeCatfig2b.ps}}}
\caption{
\footnotesize 
Selection of observed cross-correlation profiles are shown as a function of
orbital phase for HD\,81421 (left) and HD\,209295 (right).
}
\label{orbit_ccp}
\end{center}
\end{figure}

At least 15 of our 36 spectroscopic targets are binaries, giving a binarity
rate of $>$~40\%. 
We detected 7 new binaries and 11 double-lined systems.
For 9 binaries, we determined the orbit, including 4 ellipsoidal variables.
HD\,81421 is one of them, which was classified as a monoperiodic $\gamma$\,Dor star
before \citep{Martin2003A&A...401.1077M}.
However, the photometric period turns out to be half of the orbital period.
As shown on Fig.\ref{81421}, the phases of minimum light coincide with the
phases of $\gamma$ velocity at the epoch of our data.
For this object, there is no evidence for pulsations since no intrinsic
variations are observed in the cross-correlation profiles (CCPs;
Fig.\ref{orbit_ccp}, left).
For the other 5 binaries with a known orbit, none of the known photometric
periods is connected to the orbital period.
We classify them as binaries with a (candidate) $\gamma$\,Dor component.
One of them is HD\,209295, for which we confirm 4 of the known frequencies in
the cross-correlation profile variations (CPVs; Fig.\ref{orbit_ccp}, right).
For the remaining 6 binaries, our current data-set can only be used to
estimate the time-scale of the orbit.

\section{Intrinsic variations} \label{intrinsic}

\begin{figure}[]
\begin{center}
\resizebox{0.75\hsize}{!}{\rotatebox{270}{\includegraphics[clip=true]{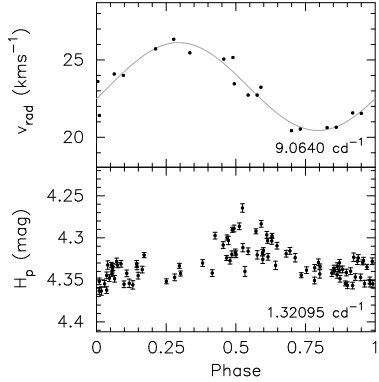}}}
\caption{
\footnotesize 
Phase diagram of the best frequency found in the radial velocity $v_{\rm rad}$ (top) and
the main frequency in the {\sc hipparcos} data (bottom) of $\gamma$\,Dor.
The reference epoch is HJD 2450000.
}
\label{27290}
\end{center}
\end{figure}

\begin{figure*}[]
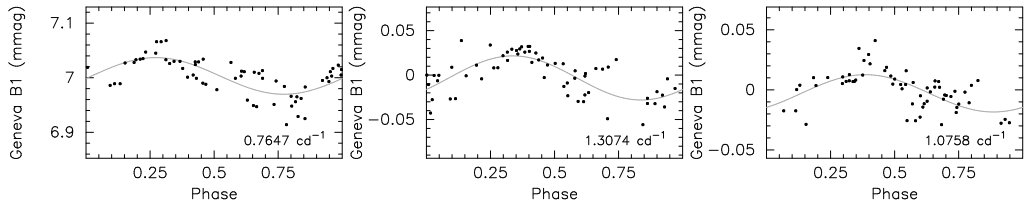

\resizebox{0.33\hsize}{!}{\rotatebox{270}{\includegraphics[clip=true]{DeCatfig4a.ps}}}
\resizebox{0.33\hsize}{!}{\rotatebox{270}{\includegraphics[clip=true]{DeCatfig4b.ps}}}
\resizebox{0.33\hsize}{!}{\rotatebox{270}{\includegraphics[clip=true]{DeCatfig4c.ps}}}
\caption{
\footnotesize 
Phase plots with  $f_1$ (left), $f_2$ (middle) and $f_3$ (right) of HD\,167858.
The reference epoch is HJD 2450000.
}
\label{intrinsicfig}
\end{figure*}

For 17 of our 36 spectroscopic targets (i.e. $\sim$45\%), CPVs are observed
(cfr. Fig.\,\ref{orbit_ccp}, right panel). 
For 8 objects, the main spectroscopic period clearly coincides with the main
period known from the {\sc hipparcos} photometry.
However, for some of them, there is no evidence for the same period in our
Johnson data. 
For 4 objects, none of the known periods from photometry is present in the
observed spectroscopic variations.
One of them is $\gamma$\,Dor itself, for which we find a $\delta$\,Scuti
period instead (Fig.\,\ref{27290}). 
Clearly, more data is needed to confirm or reject this result. 
For the remaining 5 objects showing CPVs, our current data-set is insufficient
to check the consistency between the spectroscopic and photometric variations.

Multiperiodicity is another indication of pulsation.
At least 12 of our objects show multiperiodic variations in spectroscopy
and/or photometry.
In Fig.\,\ref{intrinsicfig}, phase-diagrams with $f_1$\,=
0.7647~$\rm{c\,d^{-1}}$, $f_2$\,= 1.3074~$\rm{c\,d^{-1}}$ and $f_3$\,=
1.0758~$\rm{c\,d^{-1}}$ of HD\,167858 are given. 
$f_3$ was not known before. 

By using these criteria in combination with information found in the
literature, we upgrade 11 objects to bona-fide $\gamma$\,Dor stars.
On the other hand, 8 objects are downgraded to rejected $\gamma$\,Dor star because we
see neither intrinsic CPVs nor periodic variations in photometry (see Table\,\ref{summary}).

\section{Mode identification} \label{modeID}

\begin{figure*}[]
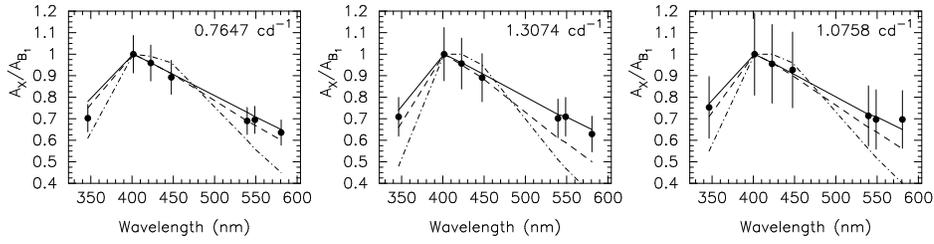

\begin{center}
\resizebox{0.3\hsize}{!}{\rotatebox{270}{\includegraphics[clip=true]{DeCatfig5a.ps}}}
\resizebox{0.3\hsize}{!}{\rotatebox{270}{\includegraphics[clip=true]{DeCatfig5b.ps}}}
\resizebox{0.3\hsize}{!}{\rotatebox{270}{\includegraphics[clip=true]{DeCatfig5c.ps}}}
\caption{
\footnotesize 
Comparison of the observed amplitude ratios (filled circles with error bars) and theoretical amplitude
ratios for $\ell$\,= 1 (full line), 2 (dashed line) and 3 (dashed-dotted line)
modes for $f_1$ (left), $f_2$ (middle) and $f_3$ (right) of HD\,167858.
}
\label{modeIDfig}
\end{center}
\end{figure*}

For the mode identification, the method of the photometric amplitude ratios is
used \citep{Dupret2003A&A...398..677D}.
For 14 objects with more than 100 Geneva datapoints, the amplitudes of the
variations with the observed frequencies were calculated.
Since the amplitudes in the $B_1$ filter are generally the biggest, we used
this filter as a reference to calculate the observed amplitude ratios and
compare them with theoretical amplitude ratios. 
The latter are obtained by performing non-adiabatic pulsation calculations with
the code {\sc mad}, which was recently upgraded to take time-dependent
convection into account (written by M.-A. Dupret), on equilibrium models
calculated with the evolution code {\sc cles} (written by R. Scuflaire).
A representative example is given in Fig.\,\ref{modeIDfig}, in which we show the
results for HD\,167858 (see example Sect.\,\ref{intrinsic}).
In this case, the observed amplitude ratios are best fitted with an $\ell$\,=
2,1 and 1 mode for $f_1$, $f_2$ and $f_3$ respectively.
Although the mode identifcation and the subsequent modelling is still ongoing
(De Ridder \& Dupret, in preparation), we only identified $\ell$\,= 1 and 2
modes so far.
 
\section{Conclusions and future prospects} \label{conclusion}

\begin{table*}
\caption{\label{summary} 
Overview of the obital and variability classification of the 11 upgraded
bona-fide and 8 downgraded rejected $\gamma$\,Dor stars, the 7 new
binaries (\underline{underlined}), the 6 binaries with new orbits (indicated
with $^{\ast}$), the 10 detected double-lined systems ({\it italics}) and the 3
suspected binaries.  
}
\begin{center}
\tabcolsep=3.5pt
\begin{tabular}{|l|l|l|l|} \hline 
        &                                    &                                                                            &                              \\[-8pt]
        &\multicolumn{1}{|l|}{bona-fide}     &\multicolumn{1}{|l|}{candidate}                                             &\multicolumn{1}{|l|}{rejected}\\[2pt] \hline 
        &                                    &                                                                            &                              \\[-8pt]
single  & HD\,14940, HD\,40745, HD\,41448,   &                                                                            &{HD\,7455, }                   \\
        & HD\,112685, HD\,135825, HD\,149989,&                                                                            &{HD\,22001,}                   \\
        & HD\,187025, HD\,216910, HD\,218225,&                                                                            &{HD\,33262 }                   \\
        & HD\,211699                         &                                                                            &{         }                   \\[2pt] \hline
        &                                    &                                                                            &                              \\[-8pt]
suspect &                                    & HD\,111829, HD\,26298                                                        &{\it HD\,27604}                \\[2pt] \hline
        &                                    &                                                                            &                              \\[-8pt]
SB1     &                                    & HD\,126516$^{\ast}$                                                         & \underline{HD\,85964}$^{\ast}$\\[2pt] \hline 
        &                                    &                                                                            &                              \\[-8pt]
SB2     &{\it \underline{HD\,34025}}$^{\ast}$&HD\,10167$^{\ast}$,{\it \underline{HD\,27377}}$^1$,{\it \underline{HD\,35416}} &{\it \underline{HD\,5590},}    \\ 
        &                                    &{\it \underline{HD\,111709}}$^{1,2}$,{\it HD\,147787},{\it HD\,214291}$^{\ast}$&{\it \underline{HD\,8393},}    \\
        &                                    &                                                                            &{\it HD\,81421}$^{\ast}$       \\[2pt] \hline 
\multicolumn{4}{c}{}\\[-8pt]
\multicolumn{4}{l}{$^1$ ellipsoidal variability instead of pulsation can not be ruled out}\\
\multicolumn{4}{l}{$^2$ shows cross-correlation profile variations but was classified as chemically peculiar star before}
\end{tabular}
\end{center}
\end{table*}

A summary of our main results is given in Table\,\ref{summary} where the stars
with a changed orbital and/or variability classification are listed.
The full results of the {\sc coralie}, {\sc modular} and {\sc p7} observations
are given by De Cat et al. (submitted to A\&A), Eyer et al. (in
preparation) and Cuypers et al. (in preparation).
Because the exploitation of dynamical information can give additional
and independent constraints on physical properties of the components,
we will give priority to binaries with a bona-fide or candidate $\gamma$\,Dor star in
future investigations.

\begin{acknowledgements}
This work is based on observations collected with the {\sc
coralie} spectrograph, {\sc modular} photometer and {\sc p7}
photometer respectively attached to the Euler Telescope (La Silla, Chile),
the 0.5-m {\sc saao} telescope (Sutherland, South-Africa), and the
Mercator telescope (La Palma, Spain). 
We are very much indebted to all the observers.
We acknowledge support from the Fund of Scientific Research (FWO) - Flanders (Belgium) through project G.0178.02.  
\end{acknowledgements}

\bibliographystyle{aa}

\end{document}